\documentclass[runningheads,a4paper]{llncs}
\usepackage{makeidx}  % allows for indexgeneration
\usepackage{graphicx}
\usepackage{soul}
\usepackage{color}
 
\begin{document}
\mainmatter              % start of the contributions
\title{Validating Intelligent Power and Energy Systems -- A Discussion of Educational Needs}
\titlerunning{Validating Intelligent Power and Energy Systems}  % abbreviated title (for running head)
\author{
	P.~Kotsampopoulos\inst{1} \and
	N.~Hatziargyriou\inst{1}  \and
	T.~I.~Strasser\inst{2} \and
	C.~Moyo\inst{2} \and
	S.~Rohjans\inst{3} \and
    C.~Steinbrink\inst{4} \and
	S.~Lehnhoff\inst{4} \and
	P.~Palensky\inst{5} \and
	A.~A.~van~der~Meer\inst{5} \and
	D.~E.~Morales~Bondy\inst{6} \and
	K.~Heussen\inst{6}
	M.~Calin\inst{7} \and
	A.~Khavari\inst{7} \and
	M.~Sosnina\inst{7} \and 
    J.~E.~Rodriguez\inst{8} \and 
    G.~M.~Burt\inst{9}
}
\authorrunning{P.~Kotsampopoulos et al.} % abbreviated author list (for running head)
\institute{
	National Technical University of Athens, Athens, Greece \and
	AIT Austrian Institute of Technology, Vienna, Austria \and
    HAW Hamburg University of Applied Sciences, Hamburg, Germany \and
	OFFIS e.V., Oldenburg, Germany \and
	Delft University of Technology, Delft, The Netherlands \and
	Technical University of Denmark, Lyngby, Denmark \and
	European Distributed Energy Resources Lab. (DERlab) e.V., Kassel, Germany \and 
    TECNALIA Research \& Innovation, Bilbao, Spain \and 
    University of Strathclyde, Glasgow, United Kingdom \\[0.15cm]
	\email{kotsa@power.ece.ntua.gr}
}
\maketitle              % typeset the title of the contribution
%
%%%%%%%%%%%%%%%%%%%%%%%%%%%%%%%%%%%%%%%%%%%%%%%%%%%%%%%%%%%%%%%%%%%%%%%%
%%%% Section: Abstract and Keywords
%%%%%%%%%%%%%%%%%%%%%%%%%%%%%%%%%%%%%%%%%%%%%%%%%%%%%%%%%%%%%%%%%%%%%%%%
%
\begin{abstract}
Traditional power systems education and training is flanked by the demand for coping with the rising complexity of energy systems, like the integration of renewable and distributed generation, communication, control and information technology. A broad understanding of these topics by the current/future researchers and engineers is becoming more and more necessary. This paper identifies educational and training needs addressing the higher complexity of intelligent energy systems. Education needs and requirements are discussed, such as the development of systems-oriented skills and cross-disciplinary learning. Education and training possibilities and necessary tools are described focusing on classroom but also on laboratory-based learning methods. In this context, experiences of using notebooks, co-simulation approaches, hardware-in-the-loop methods and remote labs experiments are discussed.
\end{abstract}
%
%%%%%%%%%%%%%%%%%%%%%%%%%%%%%%%%%%%%%%%%%%%%%%%%%%%%%%%%%%%%%%%%%%%%%%%%
%%%% Section: Introduction
%%%%%%%%%%%%%%%%%%%%%%%%%%%%%%%%%%%%%%%%%%%%%%%%%%%%%%%%%%%%%%%%%%%%%%%%
%
\section{Introduction}
\label{sec:introduction}

The rise in the extent of renewable energy systems, electrification of transportation, technology improvements to obtain a more resilient network infrastructure and challenges to have emission free and sustainable energy supply are the main drivers for the ongoing transition towards an intelligent energy system \cite{Farhangi:2010}.

With the surge of complexity due to the myriad of players and actors, the need to understand the interconnections between all energy infrastructure components increases steadily \cite{Gungor:2011}. Technically speaking there is a higher need for the integration and interaction between these actors. This includes not only the grid physics and power devices, but corresponding automation and control systems are becoming more important in the power and energy domain, in the bid to master their complexity \cite{Liserre:2010}. These needs will be addressed by ongoing developments in the area of Information and  Communication Technology (ICT). As a result of this trend a strong focus should be put on the further research and even more on educational aspects covering the ever rising complexity in energy systems. \cite{Farhangi:2010,Gungor:2011,Strasser:2017}

Education and training in the area of intelligent energy systems, especially on smart grids, can be build on communicating the concepts of modelling and simulating components \cite{Podmore:2010} to understand how they work as a system. When engineers design intelligent concepts like energy management systems, voltage control, etc. \cite{Strasser:2015}, they need to understand the paradigms of centralized and distributed control, how to use different tools, know their strength and weaknesses and how to interconnect them \cite{Martinez2011,Vournas:2004}.

The main contribution of this paper is a discussion of the challenging needs and requirements for educating current and future researchers and engineers in the domain of intelligent power and energy systems. Corresponding possibilities and necessary tools are described as well, not only focusing on the classroom but also on laboratory-based learning methods. In this context, experiences of using notebooks, co-simulation approaches, hardware-in-the-loop methods and remote labs experiments are discussed.

The following parts of this paper are organized as follows: Section~\ref{sec:intelligent_energy_systems} briefly explains challenges in intelligent power and energy systems. Education needs and requirements are discussed in the following Section~\ref{sec:needs_and_requirements}, whereas corresponding learning approaches and tools are outlined in Section~\ref{sec:education_possibilities_and_corresponding_tools}. The paper is concluded with the main findings in Section~\ref{sec:conclusions}.
%
%\cite{Farhangi2010,Kotsampopoulos2016,Strasser2014,Strasser2017}
%
%%%%%%%%%%%%%%%%%%%%%%%%%%%%%%%%%%%%%%%%%%%%%%%%%%%%%%%%%%%%%%%%%%%%%%%%
%%%% Section: Challenges in Intelligent Power and Energy Systems
%%%%%%%%%%%%%%%%%%%%%%%%%%%%%%%%%%%%%%%%%%%%%%%%%%%%%%%%%%%%%%%%%%%%%%%%
%
\section{Intelligent Power and Energy Systems}
\label{sec:intelligent_energy_systems}

\subsection{Increased Complexity trough Multi-Domain Solutions}

\begin{figure}[!htbp]
	\centering
	\includegraphics[width=1.00\columnwidth]{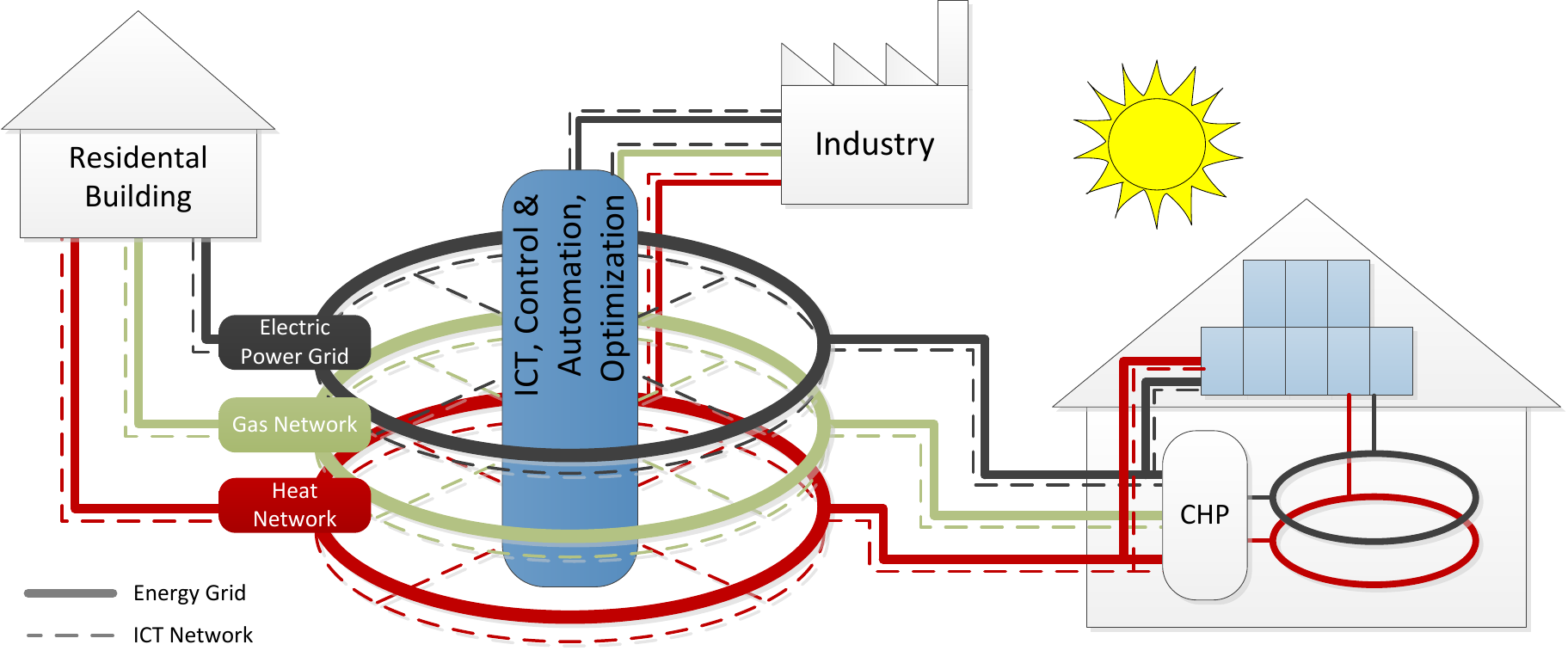}
	\caption{Illustration of intelligent energy system elements and coupling infrastructures}
	\label{fig:intelligent_energy_systems}
\end{figure}

Intelligent Power and Energy Systems can be seen as the big brother of smart grids: power systems are not only enhanced with information and communication technology, but they are integrated with other systems such as heat networks, energy markets, electric mobility, or smart city as outlined in Figure~\ref{fig:intelligent_energy_systems}. As expected, the injection of information and communication technology at a grand scale has led to the rise of systems and components that are ``intelligent'' \cite{Strasser:2015,Vrba:2014}. This has major implications on the understanding of system operations and interactions. Decentralised control strategies become more vital, in order to govern behaviour and ensure that the grid can be operated safely both within and even in ranges closer to the limits. 

Additionally, the trans-disciplinary nature of grid systems requires advanced and new types of methods for description, analysis, and optimization. Facilitating a combination of different numerical tools and domain models is the first step towards such a holistic description, but by no means is it sufficient. Non-formalized or inaccessible model information is often ``in the brains'' of experts, and requirements and constraints often remain ``in the hearts or bellies'' of stakeholders. It is crucial to put these soft factors into the loop of designing, optimizing and operating such diverse Cyber-Physical Energy systems (CPES).

\subsection{Status Quo in Education and Training}

The transformation of the traditional energy systems into a more intelligent medium opens new paths and poses new challenges. As intelligent power and energy systems present additional complexity, current and future engineers and researchers should have a broad understanding of topics of different domains, such as electric power, heat and definitely ICT related topics. Appropriate education on modern topics is essential at university level, both for undergraduate and postgraduate studies, so that future engineers will be able to understand and tackle the challenges and propose/implement new methods.

For example, the classic electrical power engineering education usually does not sufficiently cover smart grid topics, posing challenges to young researchers and students and also to the industry. Recently several universities have incorporated new courses in the undergraduate engineering curriculum or have enriched their existing courses with more modern material. Some universities have also created dedicated master courses with relevant topics. The instruction is performed with traditional methods, such as class lectures, but also with programming, advanced simulations \cite{Strasser2014} and laboratory exercises \cite{Deese:2015,Hu:2015}, where educational methods such as problem-based learning and experiential learning \cite{Kotsampopoulos2016} are applied in some cases.

Moreover, the ongoing training of current professional engineers on modern topics is important. In some cases professionals tend to be hesitant of change, and this can also be evidenced by the reluctance of industry in adopting modern solutions. By proper training, professional engineers can better understand the benefits of modern solutions and ways in which to apply them so as to improve their work. For effective training, the material should be carefully designed, for example by focusing more on the practical aspects than the theoretic background. At this point it is important to highlight that frequently power systems professionals lack thorough understanding of ICT topics. On the other hand, ICT professionals often find it hard to understand the operation of the power system. As these areas (among others) are closely connected due to the emergence of intelligent power and energy systems, it is important to create links between them. Of course, a thorough understanding of all areas (electric power, heat, ICT, automation, etc.) is difficult to achieve, however an understanding of the fundamentals of each area, without sacrificing the expert focus in each particular field, will become more and more important. The same applies to researchers who are working to find solutions beyond the state-of-the art. 
%
%%%%%%%%%%%%%%%%%%%%%%%%%%%%%%%%%%%%%%%%%%%%%%%%%%%%%%%%%%%%%%%%%%%%%%%%
%%%% Section: Learning Needs and Requirements
%%%%%%%%%%%%%%%%%%%%%%%%%%%%%%%%%%%%%%%%%%%%%%%%%%%%%%%%%%%%%%%%%%%%%%%%
%
\section{Learning Needs and Requirements}
\label{sec:needs_and_requirements}
Intelligent power and energy systems are concerned with the communicative coupling of relevant actors (producers, consumers, network operators etc.) to optimize and monitor their interconnected parts so as to result in efficient and reliable system operation. Additionally, an increasing number of decentralized, renewable energy resources (photovoltaics, wind energy, biomass, heat pumps, etc.) have to be integrated.

The problem is, however, considerably more difficult because only an integrated view of all influencing factors, such as user acceptance, CO$_2$ emissions or security of this socio-technical system, while still being held in tension by profitability, reliability, and ecological sustainability can produce practically realizable solutions that adequately take into account the complex interrelations.

For this reason, energy informatics and automation not only provides the system intelligence algorithms for adaptive control and continuous dynamic optimization of the complex and very extensive power and energy supply, but also provides the methods to create and orchestrate overall system competences (complexity control through decomposition and abstraction, identifying and focusing on general principles, finding decoupling points for effective governance, avoiding bottlenecks, etc.).

\begin{figure}[!htbp]
	\centering
	\includegraphics[width=0.60\columnwidth]{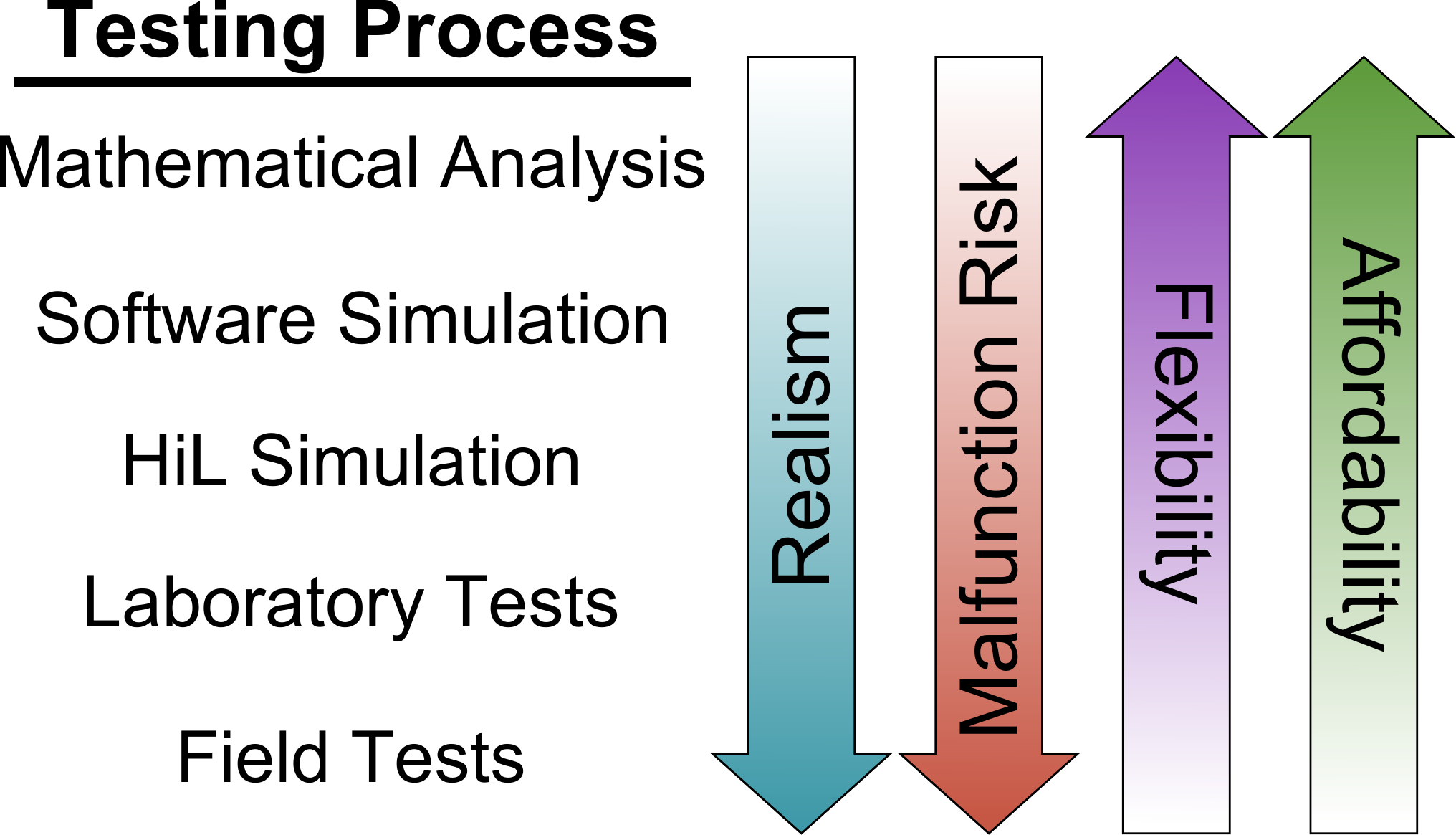}
	\caption{Steps in the testing process for smart grid components \cite{steinbrink2016}}
	\label{fig:validation}
\end{figure}

In such complex systems of systems, the design and validation is a multi-stage process, as briefly outlined in Figure~\ref{fig:validation}. Conceptual development stages already involve rough tests leading to a ``proof of concept''. When the development subject is refined, so are the tests. This allows for early identification of errors and thus makes the development process more efficient.

The earliest testing stages may often-times be realized by pure mathematical calculations. However, as soon as interactions of different components in CPES are considered, test cases become too complex for analytical approaches to be feasible. Hardware experiments, on the other hand, excel at reproducing interaction dynamics close to reality -- especially if installed in the actual energy system as field tests; however, it is important to bear in  mind the possible safety and cost consequences in the event of malfunction affecting hardware. % however, have harsh consequences in this context since hardware is expensive. In field tests the overall power system stability may even be jeopardized by poorly designed test subjects. 
Therefore, hardware setups are only suitable for the late stages of the development and testing process \cite{Strasser:2017}. 

% {kheu revisions and extension}

%The advent of the smart grid also marks the time of 
The harsh trade-offs between analytical assessments and hardware testing underline the increasing importance of software simulation for power and energy systems research and development. 
%Development and implementation is usually cheaper than hardware purchase. Computation power allows handling of more complex systems than possible with purely analytical approaches. Finally, software systems are exceptionally flexible and allow fast creation of and switching between test cases. 
Various software tools exist for simulation-based analysis of CPES. Review articles like \cite{mets2014,pochacker2013,Strasser2014} typically suggest that different tools are needed for different purposes. Also domains of expertise are wide spread, including electronics, ICT, automation, politics, economics, energy meteorology, sociology and much more. Each domain requires a specialised set of skills, but also the ability to interact and coordinate the solution of trans-disciplinary challenges. %After all, smart grid stakeholders and researchers come from fields as diverse as power electronics, ICT, automation, politics, economics, energy meteorology, sociology and many more. 

%It is a long way from the current conceptualization phase to the actual international implementation of cyber-physical energy systems. There is, however, no doubt that proper validation methods are needed for each step along this way. 
These future experts require the relevant insight and abilities to coordinate and execute design and validation of CPES solutions. These abilities are facilitated by both generic engineering and cross-disciplinary technical competences.

The generic competences include the conception, design and analysis of systems (the CDIO skills catalogue \cite{crawley2011cdio}), which can be supported, for example, by project oriented teaching methods. With increasing problem complexity, systems-oriented skills can be strengthened, such as problem decomposition, abstraction and multi-disciplinary coordination of engineering challenges. 

Cross-disciplinary learning is also required as the integration and interdependency of software and hardware systems is increased. Engineering students who aim to design and work with CPES solutions like complex control, supervisory and decision support systems, data analytics, require an increased level of programming and systems conceptual design competences, and a pragmatic view on the applicability of methods. This means that some familiarity with  domain specific system architectures and description methods is useful (references architectures like SGAM\footnote{Smart Grid Architecture Model} and RAMI\footnote{Reference Architectural Model for Industry 4.0}, use cases, test cases, etc. \cite{gottschalk2017,neureiter2014}). Basic familiarity with distributed software systems problems is also a present knowledge gap within engineering education.

Complementary in strengthening required competences in CPES education, simulation-based tools are useful for the emulation and experience of physical behaviours and cyber-physical system couplings.

Summarizing, the following needs addressing education in the domain of intelligent power and energy systems have to be addressed:

\begin{itemize}
	\item Understanding the physical behaviour of CPES and its connected sub-systems and components
	\item Understanding automation and control systems
	\item Understanding communication networks
	\item Understanding advanced control, optimization, and data analytics
\end{itemize}

Therefore, A holistic understanding of the physical and the cyber part of intelligent power and energy systems is necessary in order to design and develop a reliable and sustainable energy infrastructure of the future. 

%\comment{\textbf{Aspects of the 'practice gap'} --- \textit{can be deleted}:}
%( teaching modules typically introduce only one type of conceptual solution - how can you teach the ability to estimate the gap between 'present knowledge/maturity' and 'required knowldege/maturity'))

%new analytical methods and tools as well as design principles and programming concepts for development of IT (information technology) systems for intelligent control systems, which includes real-time analysis and reactive planning for applications in robotics, advanced process-automation, as well as energy systems and smart grids. Supporting CDIO skills development, emphasis is placed on clear presentation of ideas, concepts and knowledge throughout the project lifecycle, including and systematic evaluation of results. 
%
%%%%%%%%%%%%%%%%%%%%%%%%%%%%%%%%%%%%%%%%%%%%%%%%%%%%%%%%%%%%%%%%%%%%%%%%
%%%% Section: Training and Education Possibilities
%%%%%%%%%%%%%%%%%%%%%%%%%%%%%%%%%%%%%%%%%%%%%%%%%%%%%%%%%%%%%%%%%%%%%%%%
%
\section{Education Possibilities and Corresponding Tools}
\label{sec:education_possibilities_and_corresponding_tools}

\subsection{Classroom Education and Training}

%\textbf{[Input Esteban& Kai/DTU:]} \textit{(i) Integration of concept learning and hands-on application experience - design of “ipython notebooks”; (ii) (possible contributions to: problem-based learning / use of simulators in teaching context)}

Given the cross-disciplinary nature of intelligent power and energy systems as outlined above, students learning about this topic are exposed to a wide set of tools and concepts related to different knowledge domains. Thus, new education methods and tools must be developed, which are capable of bringing the different knowledge domains together, and allowing the students to understand the coupling and interaction of elements within intelligent solutions. 

It is established in Section~\ref{sec:needs_and_requirements} that simulations will play a role in the testing process of new solutions. It is therefore natural that students learn to use domain-specific simulation tools, both in standalone use cases and in co-simulation setups. From an educational perspective, methods that allow students to understand the coupling of theory and application (through simulation or laboratory experiments) are required.

In the following two very interesting and promising classroom examples are discussed in more detail in order to see how such courses can fulfill the challenging needs in educating about intelligent power and energy solutions. \\[-0.5em]

\noindent \textit{Example: Jupyter Notebooks \\[-0.5em]}

\begin{figure}[!htb]
	\centering
	\includegraphics[width=1.00\columnwidth]{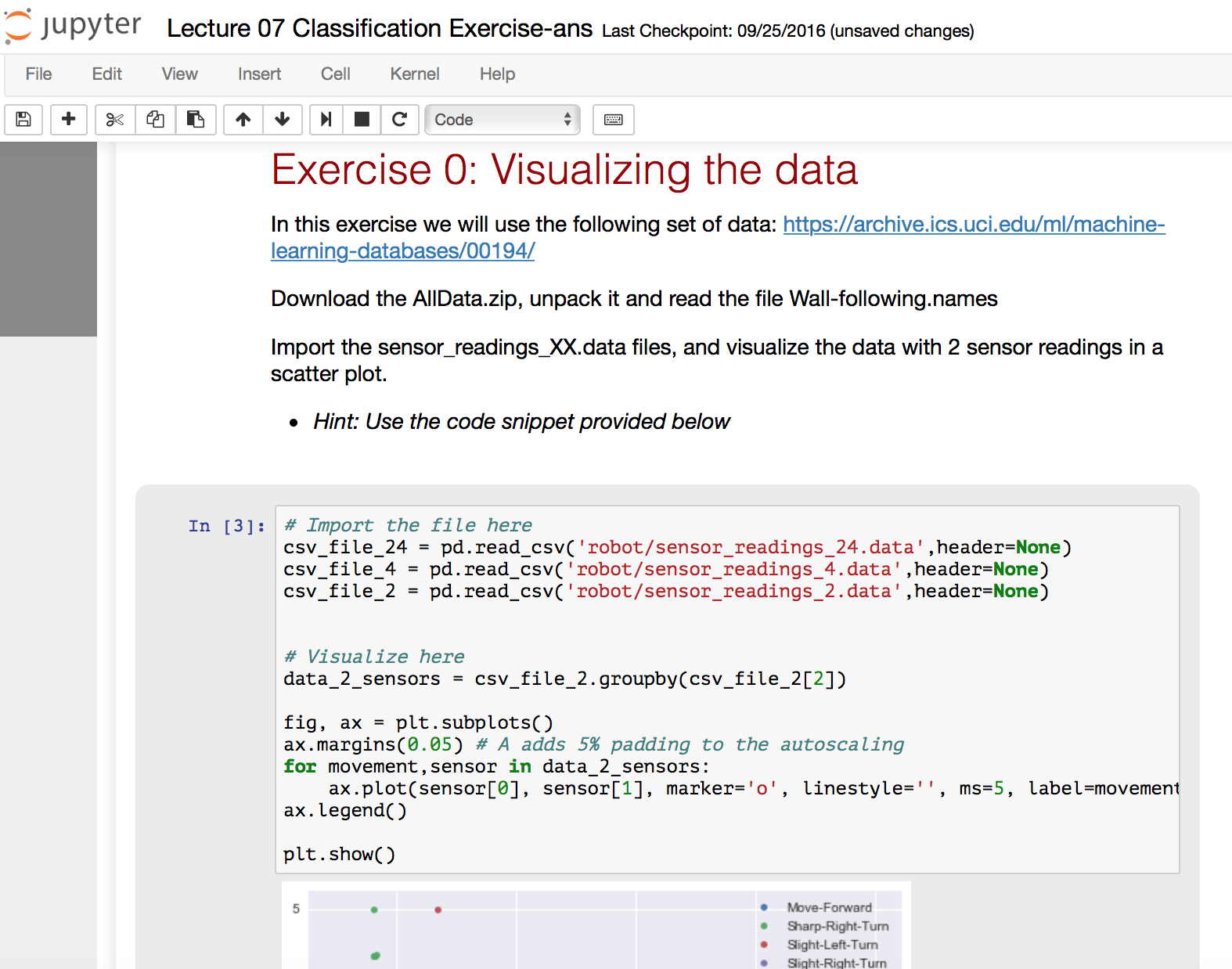}
	\caption{Example of a Jupyter notebook, where markup language and Python code can be used to showcase complex examples.}
	\label{fig:nb_example}
\end{figure}

\noindent One approach applied at the Intelligent Systems course at the Technical University of Denmark is the use of Jupyter notebooks\footnote{http://www.jupyter.org} using an IPython\cite{per-gra:2007} kernel. These code notebooks allow the execution of code cells along with explanatory text, as seen in Figure~\ref{fig:nb_example}.
 
The use of these notebooks provide a way to narrow the gap between theoretical concepts and application by setting up code examples where the student is able to able to receive a theoretical explanation followed by simulation results. Notebooks of this kind can be developed to cover a wide spectrum of intelligent energy systems concepts.

Furthermore, we believe that by constructing a framework where students can focus on a problem to solve, instead of dealing with issues related to programming, the students are able to better absorb and understand the core course concepts. So far, students have given positive feedback regarding their learning experience with these notebooks.

The gap between theoretical concepts and simulation can be addressed in the classroom, but it is also important that the students understand the limitations of solutions developed in pure simulation environments. As experiments on hardware are resource intensive, the Intelligent Systems course further employs a rapid prototyping scheme, by employing realistic software emulations of the target controls systems with a simulated back-end. This setup facilitates training on realistic automation software developments, avoiding the resource and time intensive maintenance of hardware platforms in the laboratory. In contrast to simulation-based solutions, the software developed on such platforms can be directly ported to the real hardware setup in the laboratory. \\[-0.5em]

\noindent \textit{Example: mosaik-based Co-simulation \\[-0.5em]}

\noindent Another approach is taken at the University of Oldenburg as part of the master specialization in energy informatics. In a practical course students learn to plan, execute, and analyse co-simulation-based experiments. The target audience are computer scientists, environmental modelling students and participants of the post graduate programme in renewable energies. They learn how to model controllable and flexible electrical loads and generators, as well as integrating them into smart grid scenarios along with appropriate control and optimization mechanisms. 
 
For this purpose, students should first derive and evaluate the differential equation-based models from the physical models and then transform them into discrete simulation models. Ultimately, they use the smart grid co-simulation framework ``mosaik'' (which was developed in Oldenburg) to use control and regulation mechanisms \cite{rohjans2013,schutte2011}.
 
One of the main objectives is to have students understand and engineer the function of distributed agent-based control and regulation concepts, as well as algorithms, in decentralized generators and consumers , and all way up to the operation of electrical energy systems -- in addition they analyse the requirements for real-time performance, resource utilization, robustness and flexibility.
 
The students are taught the basics in planning, execution and evaluation of simulation-based experiments. Special focus is put on the trade-off between accuracy and reliability of expected results and the necessary effort (Design of Experiments, Statistical Experimental Planning) to determine, as closely as possible, the interrelationships between influencing factors and observed target variables.
 
In terms of professional competences, the students learn to

\begin{itemize}
	\item Evaluate and derive discrete (time-stepped) models from continuous physical models,
	\item Implement the smart grid co-simulation framework ``mosaik'',
	\item Analyse distributed agent-based control concepts and algorithms for decentralized generators and consumers for the operation of electrical energy systems, with regards to the requirements of performance balancing, equipment utilization, robustness and flexibility,
	\item Design the basis for the planning, execution and evaluation of simulation-based experiments, and
	\item Recognize the importance of the trade-off between accuracy and reliability of expected results and the effort required (Design of Experiments, statistical experiment scheduling) in order to investigate the interrelations between influencing factors and observed target values with as few attempts as possible -- but as accurately as necessary.
\end{itemize}
 
\newpage 
 
Moreover, regarding methodological competences, the students learn to

\begin{itemize}
	\item Model simple controllable and flexible electrical loads and generators,
	\item Simulate appropriate control and regulation mechanisms in smart grid scenarios for electrical consumers and generators,
	\item Apply distributed agent-based control concepts and algorithms for decentralized generators and consumers to electrical energy system operation,
	\item Evaluate simulation results,
	\item Research information and methods for implementing the models, and
	\item Present their own hypotheses and check them with means of experimental planning and statistical scenario design.
\end{itemize}
 
Finally, the social- and self-competencies to be imparted are to

\begin{itemize}
	\item Apply the method of pair programming,
	\item Discuss the design decisions taken,
	\item Identify work packages and take on responsibility,
	\item Reflect on their own behaviour within the limited resource energy, and
	\item Take criticism and understand it as a proposal for the further development of their own behaviour. 
\end{itemize}

Classroom education and training (as presented in the previous two examples) allow for rapid prototyping of solutions.
Yet, the shortcomings of rapid prototyping are in the realistic emulation of the hardware behaviour. A way to bridge the gap between simulation and real-world implementation, through laboratory training, is discussed below. \\[-0.5em]

\subsection{Laboratory-based Training}

Laboratory-based training  can offer practical experience to power and energy system professionals, ICT engineers, university students and (young) researchers. Such kinds of exercises and courses have been part of the enginnering curriculum for decades \cite{Karady:2000}. However, new material and advanced training methods are necessary to deal with recent developments and new trends like the increased integration of renewables, controllable loads, etc. Therefore, laboratory courses need to be designed and applied in efficient and flexible laboratory environments.

Power and energy system engineering students usually lack the experience of working with real-world examples. For example, a realistic power system with several feeders, loads, transformers, generators is expensive and hard to build (contrary to other fields such as electric machines). Therefore, software solutions are typically used in laboratory exercises to perform power flow calculations, dynamic simulations etc. In some cases, experiments with real hardware are performed, however with low flexibility and typically for dedicated purposes. In the past, laboratory education in the power systems domain was performed with miniature generators and analogue transmission line models \cite{Karady:2000}. The situation has now changed with the wide deployment of personal computers and the development of suitable software which provides the freedom to model, flexibility and low cost. \\[-0.5em]

\noindent \textit{Example: Hardware-in-the-Loop Lab Course \\[-0.5em]}

\begin{figure}[!htbp]
	\centering
	\includegraphics[width=0.90\columnwidth]{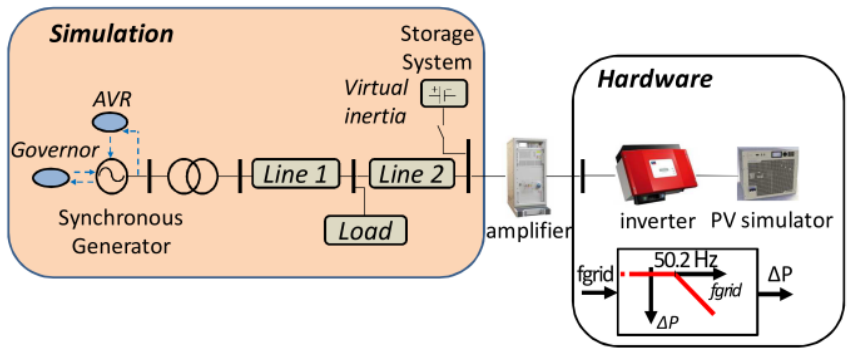}
	\caption{PHIL-based experimental setup \cite{Kotsampopoulos2016}}
	\label{fig:phil_example}
\end{figure}

Power Hardware-In-the-Loop (PHIL) simulation allows the connection of hardware devices to a real-time simulated system executed in a Digital Real-Time Simulator (DRTS). PHIL simulation presents several advantages for educational/training purposes \cite{Kotsampopoulos2016}. The real-time operation (monitor and control) is more realistic for the students. Hands-on experience of using real equipment (renewalbes, controllable loads) can be obtained, whereby actual magnitudes are measured and real equipment is controlled. Morever, equipment that is not available in the laboratory environment can be simulated, which gives freedom to design experiments (see Figure~\ref{fig:phil_example}) \cite{Kotsampopoulos2016,rohjans2013,Strasser2014}. In addition, the performance of challenging tests in a safe and controllable environment is possible. 

In this framework, laboratory exercises on modern Distributed Energy Resources (DER) related topics were designed/executed using the PHIL approach for the first time at the National Technical University of Athens \cite{Kotsampopoulos2016}. The students (both undergraduate and graduates for their dissertation) highly appreciated real-time simulation as an educational tool. Moreover, it was shown that PHIL simulation can bring the students to the hardware lab, while maintaining the modelling capability of pure simulation approaches. \\[-0.5em]

\noindent \textit{Outlook -- Improvement of Education and Training Material  \\[-0.5em]}

The above lab course example shows that the coupling of hardware equipment with a real-time simulated power grid provides various possibilities. However, additional work is necessary in order to provide a suitable and sufficient environment for teaching intelligent power and energy systems solutions. In the framework of the European project ERIGrid\footnote{https://www.erigrid.eu} the following educational activities are currently in progress to address the aforementioned requirements:

\begin{itemize}
	\item Creation of online databases with educational material,
	\item Webinars on relevant smart grid topics,
	\item Development of a remote lab for DER, and
	\item Creation of virtual labs and online software tools. 
\end{itemize}

In this context a remote lab is being developed that will allow access to the hardware of a microgrid. The users will be able to perform various experiments and corresponding tests within the grid-connected and island operation of the microgrid application. Moreover, voltage control issues (i.e., provision of ancillary services) will be studied and experiments with multi agent systems performed.
%
%%%%%%%%%%%%%%%%%%%%%%%%%%%%%%%%%%%%%%%%%%%%%%%%%%%%%%%%%%%%%%%%%%%%%%%%
%%%% Section: Conclusions
%%%%%%%%%%%%%%%%%%%%%%%%%%%%%%%%%%%%%%%%%%%%%%%%%%%%%%%%%%%%%%%%%%%%%%%%
%
\section{Conclusions}
\label{sec:conclusions}
The emergence of intelligent solutions in the domain of power and energy systems (smart grids, CPES) poses new challenges, therefore appropriate education and training approaches for students and engineers are becoming increasingly important. The trans-disciplinary nature of intelligent power and energy systems requires advanced and new types of methods for description, analysis, and optimization. A broad understanding of several areas is necessary to deal with the increased complexity and diversity. 

Challenging education needs and requirements have been identified and analysed in this study. It is explained that the validation of complex systems is a multi-stage process, while systems-oriented skills and cross-disciplinary learning needs to be cultivated. Programming and systems conceptual design competences, together with a pragmatic view are important. In order to cover the distance between theory and hands-on practice, coding and laboratory education is beneficial. The capabilities of advanced tools and methods have been discussed. The use of notebooks bridges the gap between theory and application, allowing the user to focus on solving the problem, instead of dealing with issues related to programming. Planning, executing and analysing co-simulation approaches is taught in a practical course focusing on distributed approaches. The use of PHIL simulation in laboratory education can provide hands-on experience of using real equipment in a highly flexible and controllable environment.

Future work will cover the setup of new courses, summer schools and the creation and collection of corresponding education material. 
%
%%%%%%%%%%%%%%%%%%%%%%%%%%%%%%%%%%%%%%%%%%%%%%%%%%%%%%%%%%%%%%%%%%%%%%%%
%%%% Section: Acknowledgements
%%%%%%%%%%%%%%%%%%%%%%%%%%%%%%%%%%%%%%%%%%%%%%%%%%%%%%%%%%%%%%%%%%%%%%%%
%
\subsubsection*{Acknowledgments.} This work is supported by the European Community’s Horizon 2020 Program (H2020/2014-2020) under project ``ERIGrid'' (Grant Agreement No. 654113). 
%
%%%%%%%%%%%%%%%%%%%%%%%%%%%%%%%%%%%%%%%%%%%%%%%%%%%%%%%%%%%%%%%%%%%%%%%%
%%%% Section: Literature
%%%%%%%%%%%%%%%%%%%%%%%%%%%%%%%%%%%%%%%%%%%%%%%%%%%%%%%%%%%%%%%%%%%%%%%%
%
\bibliographystyle{splncs03}
\bibliography{literature}
\end{document}